\documentstyle[12pt]{article}

\newfont{\Mb}{msbm10}

\renewcommand{\d}{{\rm d}}

\newcommand{\dx}{{\rm d}x}
\newcommand{\dy}{{\rm d}y}

\newcommand{\Pd}[2]{{{\partial #1}\over{\partial #2}}}

\begin{document}
\setcounter{equation}{0}
\setcounter{figure}{0}
\setcounter{table}{0}

\hspace\parindent
\thispagestyle{empty}

\bigskip
\bigskip
\bigskip
\begin{center}
{\LARGE \bf Finding Elementary First Integrals for}
\end{center}
\begin{center}
{\LARGE \bf  Rational Second Order Ordinary}
\end{center}
\begin{center}
{\LARGE \bf  Differential Equations}
\end{center}

\bigskip

\begin{center}
{\large
$^{a,b}$J. Avellar, $^a$L.G.S. Duarte, $^{c}$ S.E.S. Duarte and $^a$L.A.C.P. da Mota \footnote{E-mails: javellar@dft.if.uerj.br, lduarte@dft.if.uerj.br, sduarte@dft.if.uerj.br and damota@dft.if.uerj.br}
}

\end{center}

\bigskip
\centerline{\it $^a$ Universidade do Estado do Rio de Janeiro,}
\centerline{\it Instituto de F\'{\i}sica, Depto. de F\'{\i}sica Te\'orica,}
\centerline{\it 20559-900 Rio de Janeiro -- RJ, Brazil}

\bigskip
\centerline{\it $^b$ Funda\c c\~ao de Apoio \`a Escola T\'ecnica,}
\centerline{\it E.T.E. Juscelino Kubitschek,}
\centerline{\it 21311-280 Rio de Janeiro -- RJ, Brazil}

\bigskip
\centerline{\it $^c$ Funda\c c\~ao de Apoio \`a Escola T\'ecnica,}
\centerline{\it E.T.E. Visconde de Mau\'a,}
\centerline{\it 20537-200 Rio de Janeiro -- RJ, Brazil}

\bigskip
\bigskip
\abstract{Here we present an algorithm to find elementary first
integrals of rational second order ordinary differential equations
(SOODEs). In \cite{PS2}, we have presented the first algorithmic way to
deal with SOODEs, introducing the basis for the present work. In
\cite{royal}, the authors used these results and developed a method to
deal with SOODEs and a classification of those. Our present algorithm
is based on a much more solid theoretical basis (many theorems are
presented) and covers a much broader family of SOODEs than before since
we do not work with restricted ansatz. Furthermore, our present approach
allows for an easy integrability analysis of SOODEs and much faster actual calculations.}

\bigskip
\bigskip
\bigskip
\bigskip
\bigskip
\bigskip



\newpage

\section{Introduction}
\label{intro}

The differential equations (DEs) are the most widespread way to
formulate the evolution of any given system in many scientific areas.
Therefore, for the last three centuries, much effort has been made in
trying to solve them.

Broadly speaking, we may divide the approaches to solving ODEs in
the ones that classify the ODE and the ones that do not
(classificatory and non-classificatory methods).  Up to the end of
the nineteenth century, we only had many (unconnected)
classificatory methods to try to deal with the solving of ODEs.
Sophus Lie then introduced his method ~\cite{step,bluman,olver}
that was meant to be general and try to solve any ODE, i.e.,
non-classificatory. Despite this appeal, the Lie approach had a
shortcoming: namely, in order to deal with the ODE, one has to
know the symmetries of the given ODE. Unfortunately, this part of
the procedure was not algorithmic (mind you that the
classificatory approach is algorithmic by nature). So, for many
decades, the Lie method was not put to much ``practical'' use
since to ``guess'' the symmetries was considered to be as hard as
guessing the solution to the ODE itself. In \cite{nossolie1,
nossolie2}, an attempt was made to make this searching for the
symmetries to the ODE practical and, consequently, make the Lie
method more used.

So, despite all these efforts, a non-classificatory {\bf algorithmic}
approach was still missing. The first (semi) algorithmic approach applicable to
solving first order ordinary differential equations (FOODEs) was made by
M. Prelle and M. Singer \cite{PS}. The attractiveness of the PS method
lies not only in the fact that it is based on a totally different theoretical point of
view but, also in that, if the given FOODE has a solution in
terms of elementary functions, the method guarantees that this solution
will be found (though, in principle it can admittedly take an infinite
amount of time to do so). The original PS method was built around a
system of two autonomous FOODEs of the form $\dot{x} = P(x,y)$,
$\dot{y}={\cal P}(x,y)$ with $P$ and ${\cal P}$ polynomials in ${\it C}[x,y]$ or,
equivalently, the form $y'=R(x,y)$, with $R(x,y)$ a rational function of
its arguments.

The PS approach has its limitations, for instance, it deals only
with {\bf rational} FOODEs. But, since it is so powerful in many
respects, it has generated many extensions
\cite{Shtokhamer,collins,chris1,chris2,llibre,firsTHEOps1,secondTHEOps1,nossoPS1CPC}

Nevertheless, all these extensions deal only with FOODEs. In particular,
the second order ordinary differential equations (SOODEs) play a very
important role, for instance, in the physical sciences. So, with this in
mind, we have produced \cite{PS2} a PS-type approach to deal with
SOODEs. This approach dealt with SOODEs that presented
elementary\footnote{For a formal definition of elementary function, see
\cite{davenport}.} solutions (with
two elementary first integrals).

Here, we present an algorithm that comes supported by several
theoretical results. It proves to be very fast (in real applications)
and, although being not completely general, we could not find any
interesting example where it could not be used.

In section \ref{earlier}, we present the state of the art up to the
present paper. In the following section, we introduce the algorithm to
find the first integral. In section \ref{Ex}, we present some examples
of SOODEs where the algorithm is successful. Finally, we present our
conclusions and point out some directions to further our work.

\section{Earlier Results}
\label{earlier}

In the paper \cite{PS}, one can find an important result that, translated to
the case of SOODEs of the form
\begin{equation}
\label{soode}
y'' = {\frac{M(x,y,y')}{N(x,y,y')}} = \phi(x,y,y'),
\end{equation}
where $M$ and $N$ are polynomials in $(x,y,y')$\footnote{From now on, f' denotes df/dx.}, can be stated as:
\bigskip

{\bf Theorem 1: }{\it If the SOODE (\ref{soode}) has a first integral that can be written
in terms of elementary functions, then it has one of the form:
\begin{equation}
\label{eq_I}
I = w_0 + \sum_i^m c_i \ln (w_i),
\end{equation}
where $m$ is an integer and the $w's$ are algebraic functions\footnote{For a formal definition of
algebraic function, see \cite{davenport}.} of $(x,y,y')$}.
\bigskip

The integrating factor for a SOODE of the form (\ref{soode}) is defined by:

\begin{equation}
\label{Rdefinition}
R (\phi-y'') = \frac{dI(x,y,y')}{dx}
\end{equation}
where $\frac{d}{dx}$ represents the total derivative with respect to
$x$.

Bellow we will present some results and definitions (previously
presented on \cite{PS2}) that we will need. First let us remember that, on the
solutions, $dI = I_x\,dx+I_y\,dy+I_{y'}\,dy'= 0$. So, from equation
(\ref{Rdefinition}), we have:

\begin{equation}
\label{dI}
R (\phi\,dx-dy') = I_x\,dx+I_y\,dy+I_{y'}\,dy' = dI = 0.
\end{equation}
Since $y'\,dx=dy$, we have
\begin{equation}
\label{di}
R \left[(\phi+S\,y')\,dx-S\,dy-dy'\right] = dI = 0,
\end{equation}
adding the null term $S\,y'\,dx-S\,dy$, where $S$ is a function
of $(x,y,y')$. From equation (\ref{di}), we have:
\begin{eqnarray}
\label{compatcond}
I_x &=& R (\phi +  S y'),\nonumber \\
I_y &=& - R S, \\
I_{y'} &=& - R, \nonumber
\end{eqnarray}
that  must satisfy the compatibility conditions. Thus,
defining the differential operator $D$:

\begin{equation}
D \equiv \partial_{x} + y' \partial_{y} + \phi\, \partial_{y'},
\end{equation}
after a little algebra, that can be shown to be equivalent to:

\begin{eqnarray}
\label{SA1}
D[S]  & = & - \phi_y + S \phi_{y'} + S^2,\\
\label{SR1}
D[R]  & = & -R (S + \phi_{y'}),\\
\label{SA3}
R_y & = & R_{y'} S + S_{y'} R.
\end{eqnarray}
Combining~(\ref{SA1}) and~(\ref{SR1}) we obtain
\begin{equation}
\label{SR2}
D[RS]  =  -R \phi_{y}.
\end{equation}

We will now briefly explain the algorithm proposed on \cite{PS2}. In
that paper, we made the conjecture that it is always possible to have an
first integral $I$ such that $R\,S$ is a rational function of $(x,y,y')$, if
the corresponding SOODE presented an elementary solution. From equation
(\ref{SR2}), we then see that this conjecture implies that both $R$ and
$S$ are also rational functions.

Using this into equation (\ref{SA1}) and trying to solve it,
we are left with the task of solving a third degree algebraic systems of
equations on the desired coefficients of the polynomials defining the
rational function $S$. Supposing that this stage was overcome, we would
then substitute the $S$ just found in equation (\ref{SR1}) and apply the
a Prelle-Singer type approach to find $R$.

Basically, that was the novelty presented on \cite{PS2}, as far as
we know, the first semi-algorithmic approach to tackle SOODEs.

\section{The Algorithm}
\label{THEALGO}

In the present section, we will present a non-classificatory algorithm to search
elementary first integrals for SOODEs. The method is based on a Darboux
type procedure and, analogously to the Prelle-Singer \cite{PS} approach
for FOODEs, it guarantees that, for a class of SOODEs, it will eventually
find the desired first integrals. The algorithm relies on several
theoretical results that will be introduced on the next sub-section.
\subsection{Theoretical Foundations}

Let us start then by a corollary to {\bf theorem 1}
concerning $S$ and $R$.

\bigskip
{\bf Corollary 1: }{\it If a SOODE of the form (\ref{soode}) has a
first order elementary first integral then the integrating factor $R$ for such a
SOODE and the function $S$ defined in the previous section can be written as
algebraic functions of $(x,y,y')$.}

\bigskip
\bigskip
{\bf Proof:}
Using the above mentioned result by Prelle and Singer,
there is always a first integral $I=w_0 + \sum_i^m c_i \ln (w_i)$
for the SOODE. So we have, using equation (\ref{Rdefinition}),
\begin{equation}
\label{RIz}
R (\frac{M}{N}-y'') = I_x+y'I_y+y''I_{y'}\Rightarrow R = -I_{y'}
\end{equation}
where $I_u\equiv\partial_uI$. From equation (\ref{eq_I}), we have:
\begin{equation}
\label{dIdx}
I_{y'}={w_0}_{y'} + \sum_i^m c_i\frac{{w_i}_{y'}}{w_i}.
\end{equation}
Then $I_{y'}$ is an algebraic function of $(x,y,y')$ and, by equation
(\ref{RIz}), so is $R$.

From equations (\ref{compatcond}), one can see that:

\begin{equation}
S = \frac{I_y}{I_{y'}} = \frac{{w_0}_{y} + \sum_i^m c_i\frac{{w_i}_{y}}{w_i}}{{w_0}_{y'} + \sum_i^m c_i\frac{{w_i}_{y'}}{w_i}}.
\end{equation}
Therefore, $S$ is also an algebraic function of $(x,y,y')$.$\Box$

In order to produce the further mathematical results we need, we are
going to re-define the functions $R$ and $S$.

Suppose we have a SOODE of the form (\ref{soode}). We will then define the
integrating factor as:
\begin{equation}
\label{calR}
{\cal R} (M-N y'') = \frac{dI(x,y,y')}{dx} = I_x\,dx+I_y\,dy+I_{y'}\,dy'
= 0.
\end{equation}

This is equivalent to use the following re-definition:
\begin{equation}
\label{newdefR}
{\cal R} = \frac{R}{N}.
\end{equation}

Proceeding analogously to what we did with equation (\ref{dI}), we get:

\begin{equation}
\label{dii}
{\cal R} \left[(M+{\cal S}\,y')\,dx-{\cal S}\,dy-N\,dy'\right] = dI = 0,
\end{equation}
adding the null term ${\cal S}\,y'\,dx-{\cal S}\,dy$, where ${\cal S}$ is a function
of $(x,y,y')$.

This is equivalent to use the following re-definition:
\begin{equation}
\label{newdefS}
{\cal S} = S\,N.
\end{equation}

Using equations (\ref{newdefR}) and (\ref{newdefS}) into equations
(\ref{SR1}), (\ref{SA3}) and (\ref{SR2}), we get:

\begin{eqnarray}
\label{newSR1}
\frac{{\cal D}[{\cal R}]}{{\cal R}}  & = & - ({\cal S} + N_x+N_y\,y'+M_{y'}),\\
\label{third}
{\cal R}_y\,N\,-\,{\cal R}\,N_y & = & N^2\,({\cal R}_{y'}\,{\cal S} + {\cal S}_{y'}\,{\cal R}) \\
\label{newSR2}
{\cal S}\,\frac{{\cal D}[{\cal R}]}{{\cal R}}+{\cal D}[{\cal S}] & = & - (N\,M_y-M\,N_y)
\end{eqnarray}
where ${\cal D}$ is defined as ${\cal D} = N\,D$. Since $N$ is polynomial
and $R$ and $S$ are algebraic, so are ${\cal R}$ and ${\cal S}$.

Now, we will introduce some theorems that will be the basis of our
algorithm.

\bigskip
{\bf Theorem 2:} {\it Consider a SOODE of the form (\ref{soode}), that presents an
elementary first integral $I$. If ${\cal S}$ is a rational
function of $(x,y,y')$, then the integrating factor ${\cal R}$ for this SOODE
can be written as:
\begin{equation}
\label{cR}
{\cal R} = \prod_i p^{n_i}_i
\end{equation}
\noindent
where $p_i$ are irreducible polynomials in $(x,y,y')$ and $n_i$ are non-zero rational numbers.}

\bigskip
\bigskip
\noindent
To prove Theorem 2 we will need the following lemma:
\bigskip

{\bf Lemma 1:} {\it Consider a function $F$ of $(x,y,z)$.
If the differential of $F$ can be written as $dF = A \left(X\,dx + Y\,dy + Z\,dz\right)$,
where $X$, $Y$ and $Z$ are polynomial functions of $(x,y,z)$ and $A$ is an algebraic function
of $(x,y,z)$, then the first order ordinary differential equation defined as
\begin{equation}
\label{AuxiliaryODE}
\frac{dy}{dx}=-\frac{X}{Y}
\end{equation}
\noindent
where $z$ is regarded as a parameter, has $F(x,y,z)=C$ (where $C$ is a constant) as a general solution}

\bigskip
\noindent
{\bf Proof of Lemma 1:}
Consider the first order ordinary differential equation defined by (\ref{AuxiliaryODE}).
If $f$ is a function of $(x,y)$ such that
\begin{equation}
\left(Y\,\partial_x-X\,\partial_y\right) \left[f(x,y)\right] = 0,
\end{equation}
\noindent
then $f(x,y)=k$ (where $k$ is a constant) is a general solution of (\ref{AuxiliaryODE}).
Applying $\left(Y\,\partial_x - X\,\partial_y \right)$ to $F(x,y,z)$ we get $Y\,F_x - X\,F_y$.
But, by hypothesis, $F_x=A\,X$ and $F_y=A\,Y$ leading to
$$\left(Y\,\partial_x - X\,\partial_y \right) \left[F(x,y,z)\right] = Y\,A\,X - X\,A\,Y = 0.$$
\noindent
This implies that $F(x,y,z)=C$ is a general solution of (\ref{AuxiliaryODE}).$\Box$

\bigskip
\noindent
{\bf Proof of Theorem 2:}
If the hypothesis of the theorem are satisfied, ${\cal S}$ is a rational function
of $(x,y,y')$. So we can write ${\cal S}=P/Q$ where $P$ and $Q$ are polynomials.
Substituting this into equation (\ref{dii})
we get
\begin{equation}
\label{diiModified}
\frac{{\cal R}}{Q} \left[(M\,Q+P\,y')\,dx-P\,dy-N\,Q\,dy'\right] = dI = 0,
\end{equation}
\noindent
By using {\bf Lemma 1} we have that $I(x,y,y')=C$ is a general solution of the FOODE defined by
\begin{equation}
\label{AuxiliaryODExy}
\frac{dy}{dx}=\frac{M\,Q+P\,y'}{P}.
\end{equation}
\noindent
If the polynomials $(M\,Q+P\,y')$ and $P$ have a common factor we can write
$(M\,Q+P\,y')=T_1\,t_{12}\,,\,\,P=T_2\,t_{12}$ and (\ref{AuxiliaryODExy}) as
\begin{equation}
\label{AuxiliaryODExy2}
\frac{dy}{dx}=\frac{T_1}{T_2}.
\end{equation}
\noindent
Since $I(x,y,y')$ is an elementary function (by hypothesis) then,
by the theorem of Prelle and Singer\cite{PS}, there exists an integrating factor
$R_{12}$ for the FOODE (\ref{AuxiliaryODExy2}) of the form
\begin{equation}
\label{R12}
R_{12} = \prod_i f^{m_i}_i
\end{equation}
\noindent
where $f_i$ are irreducible polynomials in $(x,y)$ and $m_i$ are non-zero rational numbers.
By other side, from equation (\ref{diiModified}) we have that
\begin{equation}
\label{IxIy}
I_x = \frac{{\cal R}}{Q}\,t_{12}\,T_1\,\,;\,\,\,\,\,\,I_y = \frac{{\cal R}}{Q}\,t_{12}\,T_2,
\end{equation}
\noindent
and, so, $\frac{{\cal R}}{Q}\,t_{12}$ is an integrating factor for (\ref{AuxiliaryODExy2}).
This implies that
\begin{equation}
\label{R12R}
R_{12} = {\cal F}(I)\,\frac{{\cal R}}{Q}\,t_{12},
\end{equation}
\noindent
where ${\cal F}(I)$ is a function of the first integral $I$. From (\ref{R12R}) is easily seen
that
\begin{equation}
\label{RF}
{\cal F}(I)\,{\cal R} = R_{12}\,\frac{Q}{t_{12}}
\end{equation}
\noindent
and we may notice (see (\ref{R12})) that ${\cal F}(I)\,{\cal R}$ can be expressed as $\prod p_i^{n_i}$.
But, since ${\cal R}$ is an integrating factor for the SOODE (\ref{soode}), so is
${\overline R} \equiv {\cal F}(I)\,{\cal R}$. Finally, the reasoning we apply to the pair of variables $(x,y)$
when we make use of Lemma 1, could be carried out analogously with the pairs $(x,y')$ and $(y,y')$ as well.
So, we can conclude that there exists an integrating factor of the form (\ref{cR}).$\Box$

\bigskip
\bigskip
{\bf Theorem 3: }{\it Consider a SOODE of the form (\ref{soode}), that presents an
elementary first integral $I$. If ${\cal S}$ (defined above) is a rational
function of $(x,y,y')$ (${\cal S} = P/Q$, where $P$ and $Q$ are polynomials in $(x,y,y')$), then
$({\cal R}/Q)\,|\,{\cal D}[{\cal R}/Q]$, i.e.  ${\cal D}[{\cal R}/Q]/({\cal R}/Q)$ is a polynomial.}

\bigskip
{\bf Proof of Theorem 3:}
Substituting ${\cal S} = P/Q$ in the compatibility conditions (\ref{newSR1},\ref{newSR2}), we get:

\begin{equation}
\label{SR12}
Q\,\frac{{\cal D}[{\cal R}]}{{\cal R}}   =  - P - Q\,(N_x+N_y\,y'+M_{y'}),
\end{equation}

\begin{equation}
\label{SR22}
\frac{P}{Q}\,\frac{{\cal D}[{\cal R}]}{{\cal R}}+\frac{Q\,{\cal D}[P]-P\,{\cal D}[Q]}{Q^2} = - (N\,M_y-M\,N_y)
\end{equation}

Multiplying (\ref{SR22}) by $Q$, one gets:

\begin{equation}
P\,\frac{{\cal D}[{\cal R}]}{{\cal R}}+{\cal D}[P]-P\,\frac{{\cal D}[Q]}{Q} = - Q\,(N\,M_y-M\,N_y)
\end{equation}
and finally

\begin{equation}
\label{SR23}
P\,\left(\frac{{\cal D}[{\cal R}]}{{\cal R}}- \frac{{\cal D}[Q]}{Q}\right) =
-{\cal D}[P] - Q\,(N\,M_y-M\,N_y).
\end{equation}

If one adds $-{\cal D}[Q]$ to both sides of (\ref{SR12}), one gets:

\begin{equation}
\label{SR12new}
Q\,\left(\frac{{\cal D}[{\cal R}]}{{\cal R}}- \frac{{\cal D}[Q]}{Q}\right) =
-{\cal D}[Q] - P - Q\,(N_x+N_y\,y'+M_{y'})
\end{equation}

Re-writing (\ref{SR12new},\ref{SR23}), we have:

\begin{equation}
\label{SR12newissima}
Q\,\left(\frac{{\cal D}[{\cal R}/Q]}{{\cal R}/Q}\right) =
-{\cal D}[Q] - P - Q\,(N_x+N_y\,y'+M_{y'})
\end{equation}
and

\begin{equation}
\label{SR23newissima}
P\,\left(\frac{{\cal D}[{\cal R}/Q]}{{\cal R}/Q}\right) =
-{\cal D}[P] - Q\,(N\,M_y-M\,N_y).
\end{equation}

Since, by definition, $P, Q, M$ and $N$ are polynomial and ${\cal D}$ is
a differential operator with polynomial coefficients, the right-hand
side of both (\ref{SR12newissima},\ref{SR23newissima}) are polynomial.
Therefore, in principle, $\frac{{\cal D}[{\cal R}/Q]}{{\cal R}/Q}$ is
rational. So, let us represent it as $A/B$, where $A$ and $B$ are
polynomial that do not have any common factors. By doing that, one can
write (\ref{SR12newissima},\ref{SR23newissima}) as:
\begin{eqnarray}
\label{36}
Q\,\left(\frac{A}{B}\right) = {\cal P}_1 \\
\label{37}
P\,\left(\frac{A}{B}\right) = {\cal P}_2
\end{eqnarray}
where ${\cal P}_1=-{\cal D}[Q] - P - Q\,(N_x+N_y\,y'+M_{y'})$ and ${\cal
P}_2=-{\cal D}[P] - Q\,(N\,M_y-M\,N_y)$.

In order to satisfy (\ref{36},\ref{37})
simultaneously, since $A$ and $B$ do not have common factors, it would
be necessary that $B|Q$ and $B|P$. But, again by hypothesis, $P$ and $Q$
do not have common factors. So, one can conclude that we have $B=1$.
Therefore, $A/B = \frac{{\cal D}[{\cal R}/Q]}{{\cal R}/Q} =$
polynomial.$\Box$

\bigskip
{\bf Corollary 2: }{\it Consider a SOODE of the form (\ref{soode}), that presents an
elementary first integral $I$. If ${\cal S}$ is a rational
function of $(x,y,y')$ (${\cal S} = P/Q$, where $P$ and $Q$ are polynomials in $(x,y,y')$), then ${\cal R}/Q$ can be written as:
\begin{equation}
\label{cRQ}
{\cal R}/Q = \prod_i v^{m_i}_i
\end{equation}
\noindent
where $v_i$ are irreducible eigenpolynomials (in $(x,y,y')$) of the ${\cal D}$ operator and $m_i$ are non-zero rational numbers.}

\bigskip
{\bf Proof of Corollary 2:}
From theorem 2 we have that ${\cal R} = \prod_i p^{n_i}_i$, where $p_i$
are irreducible polynomials in $(x,y,y')$ and $n_i$ are non-zero
rational numbers. Since $Q$ is polynomial, we have $Q=\prod_j
q^{k_j}_j$, where $q_j$ are irreducible polynomials in $(x,y,y')$ and
$k_j$ are non-zero positive integers. So,
\begin{equation}
\label{cRQ2}
{\cal R}/Q = \prod_i v^{m_i}_i
\end{equation}
\noindent
where $v_i$ are irreducible polynomials (in $(x,y,y')$)
and $m_i$ are non-zero rational numbers. Since, by theorem 3, ${\cal D}[R/Q]/(R/Q)$ is
polynomial, we have:
\begin{equation}
\label{cRQ3}
\frac{{\cal D}[R/Q]}{R/Q} = \frac{{\cal D}[\prod_i v^{m_i}_i]}{\prod_i
v^{m_i}_i} = \sum m_i\,\frac{{\cal D}[v_i]}{v_i} = polynomial.
\end{equation}
\noindent
So, since the $v_i$'s are irreducible, $v_i|{\cal D}[v_i].\Box$

\subsection{The Algorithm Itself}

In this section, we will make use of the mathematics constructed above
in order to produce a ``semi-algorithm"  to deal with SOODEs of the class
defined in (\ref{soode}).

From theorem 3 we can write:
\begin{equation}
{\cal R}/Q = T = \prod_i v^{m_i}_i \Rightarrow \frac{{\cal D}[T]}{T} =
\sum m_i\,\frac{{\cal D}[v_i]}{v_i} = \sum m_i\,g_i
\end{equation}
where $v_i$ are irreducible eigenpolynomials (in $(x,y,y')$) of the
{\cal D} operator and the $g's_i$ are the corresponding cofactors.

From equations (\ref{SR12newissima},\ref{SR23newissima}), we can then
write:
\begin{equation}
\label{SR12newissimanew}
Q\,\left(\frac{{\cal D}[T]}{T}\right) = Q\,\sum m_i\,g_i =
-{\cal D}[Q] - P - Q\,(N_x+N_y\,y'+M_{y'})
\end{equation}
and
\begin{equation}
\label{SR23newissimanew}
P\,\left(\frac{{\cal D}[T]}{T}\right) = P\,\sum m_i\,g_i =
-{\cal D}[P] - Q\,(N\,M_y-M\,N_y).
\end{equation}

Equations (\ref{SR12newissimanew},\ref{SR23newissimanew}) will be the
basis of our procedure. Let us begin discussing our procedure by
talking us through the main steps before formalizing the algorithm.

As it will become clear as we go along, the first step of our procedure
is the most costly one, i.e., time consuming. In the same way as in the Prelle-Singer
procedure for FOODEs \cite{PS}, determining the eigenpolynomials and corresponding cofactors for
the ${\cal D}$ operator is a task that grows as the degree of such
polynomials grow. Our present procedure start by performing this
search. Assuming that this search succeeded and we found $v_i$ and
$g_i$ for some degree $deg$, checking (\ref{SR12newissimanew},\ref{SR23newissimanew}), we see
that it remains to be determined the following set of variables: $\{m_i,
P, Q\}$, where $m_i$ is a constant and $P$ and $Q$ are polynomials. How
to determine these? Again, by inspecting
(\ref{SR12newissimanew},\ref{SR23newissimanew}), we note that if we
construct a generic polynomial $Q$ (in $(x,y,y')$)) of some degree
$deg_Q$, we can infer the maximum degree of the polynomial $P$
($deg_P$). So, we construct such polynomials (the most general ones for
the corresponding degrees) and solve
(\ref{SR12newissimanew},\ref{SR23newissimanew}) for the coefficients of
such polynomials and for $m_i$. By doing that, we would have found $m_i$, $P$
and $Q$ and, consequently, $T = \prod_i v^{m_i}_i$. Since we have ${\cal
R}/Q = T$, we would have found $R$, the integrating factor for the SOODE
in question!

Once $R$ and $S$\footnote{Please remember that $S=P/Q$.} have been determined, we
have all the partial first derivatives of the first order differential
invariant, $I(x,y,y')$,  which is constant on the solutions (see equations (\ref{compatcond})).
This invariant can then be obtained as
\[
I(x,y,y') = \int \!R\left(\phi+S{\it y'}\right)\,\dx
-\int\!\left[RS+\Pd{}{y}\int \!R\left(\phi+Sy'\right)\dx\right]\dy-
\]
\begin{equation}
\label{III}
\int\!\left\{R+\Pd{}{y'}\left[\int\!R(\phi+Sy')\dx
-\int\!\left[RS+\Pd{}{y}\int\!R(\phi+Sy')\dx\right]\dy\right)\right\}\dy'.
\label{bigint}
\end{equation}
Basically, there are a few scenarios that can arise in the case of
SOODEs (in regard to the first order differential invariants): One can
find two pair of independent ${\cal S}$ and ${\cal R}$,leading to two
independent $I's$. In this case, we would have fully integrated the
SOODE (by solving for $y'$, using one of the invariants, and substituting
in the other to find $y$). On the other hand, it is possible that we can
only find one such invariant and, in that case, we can only reduce the
SOODE to a FOODE.

This overview above indicates the main trust of our procedure, bellow we
will present a step-by-step {\it modus operandi} of how to implement the
above scenario.

\begin{itemize}
\item Steps of the Algorithm
\begin{enumerate}
\item Set $Deg=1$.
\item For the given SOODE, we determine the eigenpolynomials and the
associated cofactors,up to degree $Deg$.
\item Set $Deg_Q=1$.
\item Construct a generic polynomial $Q$ (in $(x,y,y')$) of degree
$Deg_Q$ and a generic polynomial $P$ (in $(x,y,y')$) of degree
$Deg_P=Deg_Q+{\bf MAX}(deg_M-1,deg_N)$\footnote{Where $deg_M$ and $deg_N$ are the degree of $M$ and $N$, respectively.}.
\item Try to solve equations (\ref{SR12newissimanew},\ref{SR23newissimanew})
for the coefficients defining $Q$ and $P$ and for $m_i$.
\item If we are successful, go to step 7. In the opposite case, if $Deg_Q < 10\,Deg$, we make $Deq_Q=Deg_Q+1$ and return
to step 4. If $Deg_Q=10\,Deg$, set $Deg=Deg+1$ and return to step 2.
\item Now that we have ${\cal S}$ and ${\cal R}$ we, after checking if
that solution satisfy equation (\ref{third}), in the
affirmative case, using equation (\ref{III}), calculate the associated
first order differential invariant. If the solution does not satisfy
(\ref{third}), set $Deq_Q=Deg_Q+1$ and return to step 4 above.
\end{enumerate}
\end{itemize}

\section{Examples}
\label{Ex}

In the present section, we are going to show some examples of the usage
of our method. In the first example, we are going to point out the
improvements that have been made in comparison to the first algorithmic
approach to solving SOODEs presented on \cite{PS2}. Later we will apply
the method to another two potentially interesting physical examples found
on a very interesting paper \cite{royal} where the authors use our
results \cite{PS2} and construct a classification of SOODEs and present
different heuristics to integrate each one of them. We will comment on
that classification in the light of our present theoretical results and
issuing algorithm and make some remarks about the capability of our
method to analyze the integrability of the corresponding SOODEs.
Finally, on a more mathematical tone, we will show that the method can
solve equations previously unsolved by other powerful techniques.

\subsection{First Example}
\label{ex1}

A rich source of non-linear DEs in physics are the highly non-linear
equations of General Relativity. Einstein's equations are, of course,
in general, partial DEs, but there exist classes of space-times where
the symmetry imposed reduces these equations to ODEs in one independent
variable. One such class is that of static, spherically symmetric
space-times, which depend only on the radial variable, $r$. The metric
for a general statically spherically space-time has two free functions,
$\lambda(r)$ and $\mu(r)$ say. On imposing the condition that the matter
in the spacetime is a perfect fluid, Einstein's equations reduce to
two coupled ODEs for $\lambda(r)$ and $\mu(r)$. Specifying one of these
functions reduces the problem to solving an ODE (of first or second order)
for the other.

Following this procedure, Buchdahl~\cite{buchdahl} obtained an exact
solution for a relativistic fluid sphere by considering the so-called
isotropic metric
\[
\d{s}^2 =
  (1-f)^2(1+f)^{-2}\d{t}^2-(1+f)^4[\d{r}^2+r^2(\d{\theta}^2+\sin^2\theta\,\d{\phi}^2)]
\]
with $f=f(r)$. The field equations for $f(r)$ reduce to
\[
  ff''-3f'^2-r^{-1}ff' = 0.
\]
Changing to the notation of this paper, with $y(x)=f(r)$, we get:
\begin{equation}
\label{exemplo1_eq}
  y\,y''-3\,y'^2-(1/x)\,y\,y'=0.
\end{equation}

Let us now apply our new method (section \ref{THEALGO}) to the above equation.
First, we have to calculate the eigenpolynomials of degree=1 (and the
associated cofactors) for the differential operator associated with
equation (\ref{exemplo1_eq}) given by:
\begin{equation}
\label{exemplo1_D}
{\cal D} = x\,y\,\partial_x+x\,y\,y'\,\partial_y+ y' \left( 3\,y'x+y
\right)\partial_{y'}.
\end{equation}
These are found to be:
\begin{eqnarray}
\label{exemplo1_lambdas}
v_1 = x\,\,\,\,\, & g_1 = y \\
v_2 = y\,\,\,\,\, & g_2 = x\,y'\\
v_3 = y'\,\,\,\,\, & g_3 = 3\,y'\,x+y
\end{eqnarray}

Now, following the steps of the algorithm (given on section
\ref{THEALGO}), we can find two independent solutions:

\begin{itemize}
\item{First solution}
\begin{eqnarray}
{\cal S} & = & -\,3y'x \\
{\cal R} & = & \frac{1}{x^2\,y^4}
\end{eqnarray}
Using equation (\ref{III}), we find, for this pair of ${\cal S}$ and ${\cal R}$:
\begin{equation}
\label{Iex2}
I = {\frac {y'}{{y}^{3}x}}.
\end{equation}

\item{Second solution}
\begin{eqnarray}
{\cal S} & = & -{\frac {x \left( 2\,yx+3\,y'+3\,{x}^{2}\,y' \right) }{1+{x}^{2}}} \\
{\cal R} & = & {\frac {1+{x}^{2}}{{x}^{2}{y}^{4}}}
\end{eqnarray}
Using equation (\ref{III}), we find, for this pair of ${\cal S}$ and ${\cal R}$:
\begin{equation}
\label{Iex2}
I = {\frac {y'+{x}^{2}\,y'+y\,x}{{y}^{3}\,x}}.
\end{equation}
\end{itemize}

In comparison to the method we have presented on \cite{PS2}, we see that
our present approach has some remarkable differences: It covers a much
broader ``universe'' of SOODEs since, in the present algorithm, we do
not have restrictions imposed on ${\cal R}$ (in the algorithm presented
on \cite{PS2}, we made a conjecture that led to ${\cal R}$ being
necessarily rational). The present approach is much more theoretically
sound than the previous one, we produced a lot of theorems to support
the algorithm. From a practical point of view, this new approach is much
more efficient. For example, equation (\ref{SA1}) will generated a
system of third degree algebraic equations in the coefficients we want
to determine. On the other hand, in the algorithm introduced in section
\ref{THEALGO}, equations (\ref{SR12newissimanew},
\ref{SR23newissimanew}) will mostly generate first degree algebraic
equations since only the terms ($Q\,\sum m_i\,g_i$) and ($P\,\sum m_i\,g_i$)
produce second degree equations on the desired coefficients.

In \cite{royal}, based on our results \cite{PS2}, the authors produced
an way of finding a second independent solution for the ${\cal S}$ and ${\cal R}$
from the first one\footnote{At the time they presented their paper, they
used our results from \cite{PS2} where we introduced the theory and were
concerned with the finding of only one pair of ${\cal S}$ and ${\cal
R}$}. As it is clear from the above, our present approach finds both
pairs independently. They also classify the SOODEs they dealt with into
three types. The example we presented above (that was originally treated
in \cite{PS2}) was also used by the authors in \cite{royal} as a
representative of type I SOODEs.

In the following two sub-sections, we will introduce an example for each of
the other two types.

\subsection{Second Example}
\label{ex2}

Now, we are going to consider the Helmholtz oscillator with friction:

\begin{equation}
\label{helmholtz}
y''  +c_1\,y' + c_2\,y -\beta\, y^{2} = 0.
\end{equation}

It is an well known result that the above equation is integrable for the
choice $c_2={\frac {6}{25}}\,{c_1}^{2}$ \cite{Almendal}. As we shall see,
our method takes care of such concerns, i.e., since the algorithm
translates the solving of the SOODE into solving systems of algebraic
equations, if we consider the arbitrary constants appearing on the SOODE
($c_1,c_2,\beta$) as variables (as well as the coefficients defining the
eigenpolynomials and cofactors), the
solution of such algebraic systems will provide the regions of
integrability naturally. In order to clarify this, let us then follow our
procedure in more detail than in the previous example:

In order to follow our algorithm, the first thing to do is to determine
the eigenpolynomials (and corresponding cofactors) for the ${\cal D}$
operator:
\begin{equation}
\label{Dex2}
{\cal D} = \partial_x+y'\partial_y+ \left( -c_1\,y'-c_2\,y+\beta\,{y}^{2} \right)
\partial_{y'}
\end{equation}

The equation to be solved is:
\begin{equation}
\label{Dl}
{\cal D}[v] = g\,v
\end{equation}
where $v$ is the eigenpolynomial and $g$ is the cofactor.

It turns out that for degree 1 and 2 there is no solution to (\ref{Dl}).
A generic polynomial of degree 3 is given by:
\begin{eqnarray}
\label{eqp}
v = a_{1}+a_{2}\,{y}^{2}y'+a_{3}\,y{y'}^{2}+a_{4}\,{x}^{
3}+a_{5}\,{y}^{3}+a_{6}\,{y'}^{3}+a_{7}\,xyy'+a_{8}
\,x+\nonumber \\
a_{9}\,y+a_{10}\,{x}^{2}+a_{11}\,yy'+a_{12}\,{y
}^{2}+a_{13}\,xy'+a_{14}\,xy+a_{15}\,{y'}^{2}+\nonumber \\
a_{16}\,{x}^{2}y'+a_{17}\,{x}^{2}y+a_{18}\,x{y}^{2}+a_{19}\,x{
y'}^{2}+a_{20}\,y'
\end{eqnarray}
Analyzing the ${\cal D}$ operator (eq. (\ref{Dex2})) and equation
(\ref{Dl}), one may conclude that the maximum degree for $g$ is 1.
So,
\begin{equation}
\label{lambdaex2}
g = b_1+b_2\,x+b_3\,y+b_4\,y'.
\end{equation}
Solving equation (\ref{Dl}) will be translated into solving the
following algebraic system:
\begin{eqnarray}
sys = (\beta\,a_{2}-a_{5}\,b_{3}=0,\beta\,a_{13}-{\it
c_2}\,a_{7}-a_{12}\,b_{2}-a_{18}\,b_{1}-
a_{14}\,b_{3}=0,\nonumber \\
-a_{5}\,b_{2}-a_{18}\,b_{3}+\beta\,a_{7}=0,2\,\beta\,a_{19}-a_{7}\,b_{3}
-a_{2}\,b_{2}-a_{18}\,b_{4}=0,\nonumber \\
-a_{10}\,b_{4}-a_{16}\,b_{1}-{\it
c_1}\,a_{16}-a_{13}\,b_{2}+a_{17}=0,-a_{19}\,b_{2}-a_{16}\,b_{4}=0,\nonumber \\
-a_{16}\,b_{3}-a_{17}\,b_{4}-a_{7}\,b_{2}=0,\nonumber \\
-a_{10}\,b_{3}-a_{17}\,b_{1}-a_{14}\, b_{2}-{\it
c_2}\,a_{16}=0,-a_{10}\,b_{1}-a_{8}\,b_{2}+3\,a_{4}=0,\nonumber \\
-a_{4}\,b_{2}=0,-a_{17}\,b_{2}-a_{4}\,b_{3}=0,-a_{1}\,b_{2}-a_{8}
\,b_{1}+2\,a_{10}=0,\nonumber \\
-a_{10}\,b_{2}-a_{4}\,b_{1}=0,-a_{16}\,b_{2}-a_{4}\,b_{4}=0,-a_{6}\,b_{3}-a_{3}\,b_{4}=0,\nonumber \\
-a_{18}\,b_{2}+\beta\,a_{16}-a_{17}\,b_{3}=0,-a_{3}\,b_{2}-a_{7}\,b_{4}-a_{19}\,b_{3}=0,\nonumber \\
-a_{3}\,b_{3}+3\,\beta\,a_{6}-a_{2}\,b_{4}=0,-a_{9}\,b_{3 }-{\it
c_2}\,a_{11}+a_{18}-a_{12}\,b_{1}+\beta\,a_{20}=0,\nonumber \\
-{\it c_2}\,a_{2}+\beta\,a_{11}-a_{12}\,b_{3}-a_{5}\,b_{1}=0,\nonumber \\
-{\it c_2}\,a_{20}+a_{14}- a_{1}\,b_{3}-a_{9}\,b_{1}=0,-3\,{\it c_1}\,a_{6}+a_{3}-a_{15}\,b_{4}-a_{6}\,b_{1}=0,\nonumber \\
2\,\beta\,a_{3}-a_{2}\,b_{3}-a_{5}\,b_{4}=0,a_{19}-2\,{\it
c_1}\,a_{15}+a_{11}-a_{20}\,b_{4
}-a_{15}\,b_{1}=0,\nonumber \\
-a_{6}\,b_{4}=0,a_{13}-a_{1}\,b_{4}-a_{20}\,b_{1}-{\it
c_1}\,a_{20}+a_{9}=0,\nonumber \\
a_{8}-a_{1}\,b_{1}=0,-3\,{\it
c_2}\,a_{6}+2\,a_{2}-2\,{\it
c_1}\,a_{3}-a_{15}\,b_{3}-a_{11}\,b_{4}-a_{3}\,b_{1}=0,\nonumber \\
-2\,{\it c_2}\,a_{15}+2\,a_{12}-a_{20}\,b_{3}+a_{7}-a_{9}\,b_{4}-a_{11}\,b_{1}-
{\it c_1}\,a_{11}=0,\nonumber \\
-2\,{\it c_1}\,a_{19}-a_{13}\,b_{4}-a_{19}\,b_{1}-a_{15}\,b_{2}+a_{7}=0,\nonumber \\
-a_{20}\,b_{2}-a_{8}\,b_{4}-a_{13}\,b_{1}+2\,a_{16}-{\it c_1}\,a_{13}
+a_{14}=0,-a_{6}\,b_{2}-a_{19}\,b_{4}=0,\nonumber \\
-2\,{ \it c_2}\,a_{19}-a_{11}\,b_{2}-{\it c_1}\,a_{7}-a_{13}\,b_{3}+2\,a_{18}-a_{14}\,b_{4}-a_{7}\,b_{1}=0,\nonumber \\
2\,a_{17}-{\it c_2}\,a_{13}-a_{9}\,b_{2}-a_{14}\,b_{1}-a_{8}\,b_{3}=0,\nonumber \\
-a_{12}\,b_{4}-a_{2}\,b_{1}-a_{11}\,b_{3}-{\it c_1} \,a_{2}-2\,{\it
c_2}\,a_{3}+3\,a_{5}+2\,\beta\,a_{15}=0)
\end{eqnarray}
A solution to the system above is:
\begin{eqnarray}
a_{20}=0,a_{4}=-\frac{2}{3}\,\beta\,a_{11},a_{9}={\frac {4}{25}}\,{{\it
c_1}}^{2}a_{11},{\it c_2}={\frac {6}{25}}\,{{\it
c_1}}^{2},a_{3}=0,\nonumber \\
a_{18}=0,a_{12}=0,a_{5}=0,a_{15}=0,a_{16}=0,b_{3}=0,b_{4}=0,a_{19}=0,\nonumber \\
a_{7}=0,a_{2}=0,a_{8}=0,b_{2}=0,a_{1}=0,b_{1}=-\frac{6}{5}\,
{\it c_1},a_{11}=a_{11},a_{13}=0,\nonumber \\
a_{10}=\frac{4}{5}\,{\it c_1}\,a_{11},a_{17}=0,a_{14}=0,a_{6}=0.
\end{eqnarray}
Leading to:
\begin{eqnarray}
v &=&-\frac{2}{3}\,\beta\,{y}^{3}+{\frac {4}{25}}\,{{\it
c_1}}^{2}\,{y}^{2}+\frac{4}{5}\,{\it c_1}\,y\,y'+{y'}^{2}
\nonumber \\
g&=&{-\frac{6}{5}}\,{\it c_1}\nonumber \\
{\it c_2}&=&{\frac {6}{25}}\,{{\it c_1}}^{2}
\end{eqnarray}

As we have mentioned, the solution via our procedure takes care of the
analysis of integrability of the SOODE. The above value for $c_2$ is an
well known result for the integrability of the Helmholtz system.

So, following the remaining steps of our algorithm, solving equations
(\ref{SR12newissimanew}) and (\ref{SR23newissimanew}) to find $P$ and $Q$, we can get two independent solutions:
\begin{itemize}
\item{First solution}
\begin{eqnarray}
{\cal S} & = & {\frac {-25\,\beta\,{y}^{2}+4\,{{\it c_1}}^{2}y+10\,{\it c_1}\,y'}{10
\,{\it c_1}\,y+25\,y'}} \\
{\cal R} & = & {\frac {2\,{\it c_1}\,y+5\,y'}{-50\,\beta\,{y}^{3}+12\,{{\it c_1}}^{2
}{y}^{2}+60\,{\it c_1}\,y\,y'+75\,{y'}^{2}}}
\end{eqnarray}
Using equation (\ref{III}), we find, for this pair of ${\cal S}$ and ${\cal R}$:
\begin{equation}
\label{Iex2}
I = 6\,{\it c_1}\,x+5\,\ln  \left( -12\,{{\it c_1}}^{2}{y}^{2}
+50\,\beta\,{y}^{3}-60\,{\it c_1}\,y\,y'-75\,{y'}^{2} \right)
\end{equation}

\item{Second solution}
\begin{eqnarray}
{\cal S} & = & {\frac {6\,{{\it c_1}}^{2}y+25\,{\it c_1}\,y'-25\,\beta\,{y}^{2}}{25\,y'
}} \\
{\cal R} & = & y' \left( -\frac{2}{3}\,\beta\,{y}^{3}+{\frac {4}{25}}\,{{\it c_1}}^{2}{y}^{2}+
\frac{4}{5}\,{\it c_1}\,y\,y'+{y'}^{2} \right) ^{-5/6}
\end{eqnarray}
\end{itemize}

Using equation (\ref{III}), we find, for this pair of ${\cal S}$ and ${\cal R}$:
\begin{equation}
\label{Iex2}
I =\int \frac{\! \left( 6\,{{\it c_1}}^{2}y+25\,{\it c_1}\,y'-25\,\beta\,{y}^{2}
 \right)}{\left( {\frac {4}{25}}\,{{\it c_1}}^{2}{y}^{2}-\frac{2}{3}\,\beta\,{y
}^{3}+\frac{4}{5}\,{\it c_1}\,y\,y'+{y'}^{2} \right) ^{-5/6}}{dy}
\end{equation}

What new properties of our algorithm can be emphasized by this example?
As mentioned, the method is blind as for the origin of the parameters we
want to determine. So, one can use this to include the integrability
analysis in a very natural way.  We see that, since our method is based on
very sound theoretical results (see section \ref{THEALGO}) about the nature of
${\cal R}$, we are left with a truly algorithmic procedure of the PS-
type, i.e., we reduce the solving of a SOODE to an algebraic problem.

\subsection{Third Example}
\label{ex3}

The next example is also physically motivated, the force free Duffing-van
der Pol oscillator:
\begin{equation}
\label{exemplo3_eq}
y'' + \left( \alpha+\beta\,y^{2} \right)\,y'-\gamma\,y+y^{3}=0
\end{equation}

For this case, there are also some integrability aspects. We are not
going to dwell on that now. This was already done on the last example.
It is well known (see for instance, \cite{LakshmananRajasekar}) that the integrability occurs for ${\it
\gamma}=-3\,{\beta}^{-2},\alpha=4\,{\beta}^{-1}$.

Applying our method (section \ref{THEALGO}) to that equation, we first
have to calculate the eigenpolynomials to the appropriate degree (and the associated
cofactors) for the differential operator associated with equation
(\ref{exemplo3_eq}) given by:
\begin{equation}
\label{exemplo3_D}
{\cal D} = \partial_x+y'\partial_y+ \left( -y'\,\alpha-
\beta\,{y}^{2}\,y'+\gamma\,y-{y}^{3} \right) \partial_{y'}
\end{equation}
These are:
\begin{eqnarray}
\label{exemplo3_lambdas}
v_1 = {\frac {\left( y+y'\beta \right) }{\beta}}\,\,\,\,\,\,\,\,\,\,\,g_1 = - {\frac {3+{\beta}^{2}{y}^{2}}{\beta}} \\
v_2 = {\frac {\left( 3\,y+{\beta}^{2}{y}^{3}+3\,y'\beta \right) }{3\,\beta}}\,\,\,\,\,\,\,\,g_2 = -\frac{3}{\beta}
\end{eqnarray}

Now, following the steps of the algorithm (given on section
\ref{THEALGO}), we can find two independent solutions:

\begin{itemize}
\item{First solution}
\begin{eqnarray}
{\cal S} & = & {\frac {{\beta}^{2}{y}^{2}+1}{\beta}}\\
{\cal R} & = & {\frac {3\,\beta}{\left( 3\,y+{\beta}^{2}{y}^{3}+3\,y'\beta \right) }}
\end{eqnarray}

Using equation (\ref{III}), we find, for this pair of ${\cal S}$ and ${\cal R}$:
\begin{equation}
\label{Iex2}
I ={\frac {3\,x+\ln  \left( 3\,y\beta\,\alpha-9\,y+{\beta}^{2}{y}^{3}+3\,
y'\beta \right) \beta}{\beta}}
\end{equation}

\item{Second solution}
\begin{eqnarray}
{\cal S} & = &{\frac {4\,y'\beta+y'{\beta}^{3}{y}^{2}+3\,y+{y}^{3}{\beta}^{2}}{y'}} \\
{\cal R} & = &{\frac {y'}{\sqrt [3]{y+\frac{1}{3}\,{y}^{3}{\beta}^{2}+y'\beta} \left( y'\beta+y
 \right) }}
\end{eqnarray}
\end{itemize}

Using equation (\ref{III}), we find, for this pair of ${\cal S}$ and ${\cal R}$:
\begin{eqnarray}
\label{Iex2}
I = - \left( 3\,y+{y}^{3}{\beta}^{2}+3\,y'\beta \right) ^{2/3}
{\beta}^{2/3}+2\,\ln  \left( \sqrt [3]{3\,y+{y}^{3}{\beta}^{2}+3\,y'
\beta}-y\,{\beta}^{2/3} \right) - \nonumber\\
\ln  \left(  \left( 3\,y+{y}^{3}{\beta}
^{2}+3\,y'\beta \right) ^{2/3}+\sqrt [3]{3\,y+{y}^{3}{\beta}^{2}+3\,y'
\beta}\,y\,{\beta}^{2/3}+{y}^{2}{\beta}^{4/3} \right) + \nonumber \\
2\,\sqrt {3}\arctan
 \left( {\frac {\sqrt {3} \left( 2\,\sqrt [3]{3\,y+{y}^{3}{\beta}
^{2}+3\,y'\beta}+y\,{\beta}^{2/3} \right) }{3\,y\,{\beta}^{2/3}}} \right)
\end{eqnarray}

In this example, we skipped the integrability analysis (that was already
exemplified on the previous one but could be done in the same fashion
here) and followed the same procedure as in the two first examples to
find the functions ${\cal S}$ and ${\cal R}$.

\subsection{Some Comments on the Examples Above}
\label{commentsEX}

Before we present the fourth (and final) example, let us make some
considerations on the three previous ones: In \cite{PS2}, we have
introduced the basis for an approach to algorithmically solve SOODEs.
Later on, \cite{royal} picked up some of these ideas and have developed
a different approach to the task. They also used our function $S$ as one
of the main ingredients of their method. Shortly, they say they use the
same ansatz as we did in \cite{PS2} for $S$:
\begin{equation}
\label{Sroyal}
S = \frac{a(x,y)+b(x,y)\,y'}{c(x,y)+d(x,y)\,y'}
\end{equation}
Actually, this is a more general form than the one we considered in
\cite{PS2}. As we do here, we consider $S$ to be a plain rational
function on $(x,y,y')$ (not rational in respect only to $y'$). Using
their format (\ref{Sroyal}), our case would correspond to $(a,b,c,d)$
being polynomials on $(x,y)$\footnote{Please note that, in our considerations,
we were not limited to numerators and denominators of degree 1 in y',
etc.}. This is an important point in understanding the differences
between our present algorithmic method and the approach presented on
\cite{royal}. To further our discussion, let us introduce the results
for $S$ ad $R$, presented on \cite{royal}, for the three examples above:

\begin{itemize}

\item  For example 1 above (\ref{ex1})

The authors of \cite{royal} have found two pairs of $S$ and $R$:
\begin{eqnarray}
\label{royalex1}
S_1 =-{\frac{3\,y'}{x}}\,\,\,\,  R_1 = {\frac{1}{y^3\,x}}\nonumber \\
S_2 =-{\frac{y'}{x}} \,\,\,\, R_2 = {\frac{1}{y^5\,x}}
\end{eqnarray}

\item  For example 2 above (\ref{ex2})

For this case, the authors have found:
\begin{eqnarray}
\label{royalex2}
S_1 = {\frac {-
\beta\,{y}^{2}+\frac{4\,c_1^2}{25}\,y+\frac{2\,c_1}{5}\,y'}{\frac{2\,c_1}{5}
\,y+y'}} &&  R_1 = -(y'+\frac{2\,c_1\,y}{5})\,e^{\frac{6\,c_1}{5}x}\nonumber \\
S_2 ={\frac {-
\beta\,{y}^{2}+\frac{6\,c_1^2}{25}\,y+\frac{c_1}{5}\,y'}{y'}} && R_2 = -y'\,e^{c_1\,x}
\end{eqnarray}

\item  For example 3 above (\ref{ex3})

In this third example, the authors of \cite{royal} have found only one pair of $S$ and $R$:
\begin{eqnarray}
\label{royalex3}
S_1 = \frac{1}{\beta}+\beta\,y^2 &&  R_1 =  e^{\frac{3\,x}{\beta}}
\end{eqnarray}

\end{itemize}

In order to obtain the above results, the authors also used an ansatz
for $R$:
\begin{equation}
R = A(x,y) + B(x,y)\,y'.
\end{equation}

They use the two ansatz (for $S$ and $R$) into equations (\ref{SA1}),
(\ref{SR1}) and (\ref{SA3}) thus obtaining a system of partial
differential equations for the unknown functions ($a$,$b$,$c$,$d$, $A$,$B$) of
$(x,y)$, this can be very trick and not algorithmic in essence (see
example 3 above where the authors could only find one independent pair
of $S$ and $R$). On the other hand, our method only consider that $S$
has to be a rational function of $(x,y,y')$ and, due to the theoretical
results presented on section \ref{THEALGO}, we know the general form of $R$
and that led to an algorithmic Darboux type procedure to determine $S$
and $R$.

As can be seen from equations (\ref{royalex2}, \ref{royalex3}), the
method presented on \cite{royal} can find, in principle, non-algebraic
$S$ and $R$ (our algorithm deals with rational $S$ and algebraic $R$).
But, for the examples presented on \cite{royal}, the authors found only
rational $S$ functions and, therefore, in principle, inside the
applicability of our method. As mentioned already, the point is that our
approach is algorithmic (in the Prelle-Singer sense) and, computationally
speaking, much faster to apply. Basically, we convert solving a SOODE to solving an
algebraic system while the method presented on \cite{royal} transforms
the solving of the SOODE in solving a system of partial differential
equations.

The three examples above are physically motivated examples and were
previously dealt with in \cite{PS2} (example 1) and \cite{royal}
(examples 1,2,3). In this latter reference, these were classified into
three different types. It is worth to point out that, in our algorithm,
that is not necessary, i.e., all examples are treated equally since our
procedure is non-classificatory.

In order to conclude this section with examples, we will pass now to a
more mathematical note and present an academic SOODE with the intention
to show that the present approach is capable of solving it while many
powerful techniques fail.

\subsection{Fourth Example}
\label{ex4}

Let us consider the following SOODE:
\begin{equation}
\label{exemplo4_eq}
y'' = -{\frac {{y'}^{2}}{-y-1+3\,x\,y'}}
\end{equation}

Let us now apply our new method (section \ref{THEALGO}) to the above equation.
First, we have to calculate the eigenpolynomials of degree=1 (and the
associated cofactors) for the differential operator associated with
equation (\ref{exemplo1_eq}) given by:
\begin{equation}
\label{exemplo1_D}
{\cal D} = \left( -y-1+3\,x\,y' \right) \partial_x +
\left( -y'y-y'+3\,{y'}^{2}x \right) \partial_y - {y'}^{2} \partial_{y'}.
\end{equation}
These are found to be (up to degree=1):
\begin{eqnarray}
\label{exemplo1_lambdas}
v_1 = y'\,\,\,\,\, & g_1 = -y'
\end{eqnarray}

Now, following the steps of the algorithm (given on section
\ref{THEALGO}), we can find two independent solutions:

\begin{itemize}
\item{First solution}
\begin{eqnarray}
{\cal S} & = & -{\frac {y' \left( -2\,y+3\,x\,y'-2 \right) }{ \left( -y-1+3\,x\,y' \right)}} \\
{\cal R} & = & -y-1+3\,x\,y'
\end{eqnarray}
Using equation (\ref{III}), we find, for this pair of ${\cal S}$ and ${\cal R}$:
\begin{equation}
\label{Iex41}
I =2\,x\,{y'}^{2}y+2\,{y'}^{2}x-2\,{y'}^{3}{x}^{2}-\frac{2}{3}\,{y}^{2}y'-\frac{4}{3}\,y\,y'-\frac{2}{3}\,
y'.
\end{equation}

\item{Second solution}
\begin{eqnarray}
{\cal S} & = &  \left( -\frac{1}{2}+\frac{i\sqrt {3}}{2} \right) y'\\
{\cal R} & = & {y'}^{( -\frac{1}{2}+\frac{i\sqrt {3}}{2} )}
\end{eqnarray}
Using equation (\ref{III}), we find, for this pair of ${\cal S}$ and ${\cal R}$:
\begin{equation}
\label{Iex42}
I = \arcsin \left( {\frac {y\sqrt {3}+\sqrt {3}}{\sqrt { \left( 6\,xy'-3\,y
-3 \right) ^{2}+ \left( y\sqrt {3}+\sqrt {3} \right) ^{2}}}} \right) +
\frac{\sqrt {3}}{2}\,\ln  \left( y' \right).
\end{equation}
\end{itemize}

The great advantage of having a non-classificatory, algorithmic approach
is that we can tackle every SOODE in the same manner. The above example,
despite its simple appearance, could not even be reduced by the powerful
Maple solver (release 9.5). In our method, the most costly part of the
algorithm is to determine the eigenpolynomials (and corresponding cofactors) and, for this example, this is actually very simple.

\section{Conclusion}
\label{conclu}

In \cite{PS2}, we have developed a method, based on a conjecture,
to deal with SOODEs that presented an elementary solution (possessing
two elementary first integrals).

In that same paper, we have introduced a function $S$ to transform the
Pfaffian equation related to the particular SOODE under consideration
into a 1-form proportional to the differential of the first integral.
That function $S$ was instrumental in finding the integrating factor for
the SOODE.

Here, in the present paper, we introduce many theoretical results
concerning that function $S$ and present an way to calculated it and the
integrating factor $R$ via a Darboux-type procedure.

Briefly, we construct a differential operator ${\cal D}$,
\begin{equation}
{\cal D} = N\,D = N\,\partial_{x} + y'\,N \partial_{y} + M\, \partial_{y'},
\end{equation}
extracted from the SOODE, and the corresponding eigenpolynomials and
cofactors will be the building blocks of $S$ and $R$. Furthermore, the
procedure is semi- algorithmic and, given enough time, if the solution
exists, it will find it (as does the Prelle-Singer approach for FOODEs).

From an operational point of view, our method is very sound, since, as
mentioned in sub-section \ref{ex2}, our algorithm converts the solving of
the SOODE into solving (essentially) first degree algebraic equations.

As a consequence of this, our approach is capable of analyzing the
integrability regions for the SOODE (for the case where it presents
undetermined parameters). Since we translate the SOODE into an algebraic
system, if we consider those parameters as variables, the method solves
the system given the values for which there is an integration possible.

Our method is not general since it is limited to the cases where there
exists an elementary first integral and, furthermore, to the cases where
$S$ is rational. Actually, in respect to the first limitation mentioned
above, it is possible that our method solves a case where the first
order differential invariant is not elementary (see example 2, sub-section
\ref{ex2}). The point is that, if $S$ and $R$ are of the right format
(i.e., rational $S$ and $R=\prod_i p_i^{n_i}$, where $p_i$ are
irreducible polynomials in $(x,y,y')$ and $n_i$ are non-zero rational
numbers.), our method will work. The method is sure to work if there is
an elementary first order invariant. We would like to elaborate a little
bit bout the second restriction mentioned above ($S$ being rational):
although all of our theoretical results apply to that particular
situation, we could not, so far, find any an interesting (not easily
solvable) example where $S$ is an algebraic function of $(x,y,y')$ (the
general case for elementary first order invariants, see Corollary 1
above). So, the restriction does not seem to be very restrictive. For
instance, in \cite{royal}, all cases of physical interest are covered by
our method, i.e., present rational $S$.

In regard to future work, many extensions of the present paper can be
pursued: We intend to further our work to include Liouvillian first
order differential invariants (along the lines we followed, in dealing
with FOODEs \cite{secondTHEOps1}). We can also include SOODEs with
elementary functions (see \cite{Shtokhamer,nossoPS1CPC}, for related work
applied to FOODEs). As hinted above, we also intend to further analyze the
capabilities of our method in the determination of integrability regions
for the parameters present on the SOODE.



\begin{thebibliography}{25}

\bibitem{step} Stephani, H. {\it Differential equations: their
solution using symmetries}, ed.\ M.A.H. MacCallum, Cambridge University
Press, New York and London (1989).

\bibitem{bluman} G.W. Bluman and S. Kumei, {\it Symmetries and Differential
Equations}, Applied Mathematical Sciences {\bf 81}, Springer-Verlag, (1989).

\bibitem{olver} P.J. Olver, {\it Applications of Lie Groups to Differential
Equations}, Springer-Verlag, (1986).

\bibitem{nossolie1} E.S. Cheb-Terrab, L.G.S. Duarte and L.A.C.P. da Mota, {\it
Computer Algebra Solving of First Order ODEs Using Symmetry Methods}.
{\it Comput.Phys.Commun.}, {\bf 101}, (1997), 254.

\bibitem{nossolie2} E.S. Cheb-Terrab, L.G.S. Duarte and L.A.C.P. da Mota, {\it
Computer Algebra Solving of Second Order ODEs Using Symmetry Methods}.
{\it Comput.Phys.Commun.}, {\bf 108}, (1998), 90.

\bibitem{PS}
M Prelle and M Singer,
Elementary first integral of differential equations.
{\it Trans. Amer. Math. Soc.}, {\bf  279} 215 (1983).

\bibitem{Shtokhamer} R. Shtokhamer, {\it Solving first order differential
equations using the Prelle-Singer algorithm}, Technical report 88-09, Center for
Mathematical Computation, University of Delaware (1988).

\bibitem{collins}
C B Collins,
Algebraic Invariants Curves of Polynomial Vector Fields in the Plane,
{\it Preprint.} Canada: University of Waterloo (1993); C B Collins,
Quadratic Vector Fields Possessing a Centre,
{\it Preprint.} Canada: University of Waterloo (1993).

\bibitem{chris1} C. Christopher, Liouvillian first integrals of
second order polynomial differential equations. {\it Electron. J.
Differential Equations} 1999, No. 49, 7 pp. (electronic).

\bibitem{chris2} C. Christopher and J. Llibre, Integrability via
invariant algebraic curves for Planar polynomial differential systems
{\it Ann.Differential Equations} {\bf 16} (2000), no. 1, 5-19.

\bibitem{llibre} J. Llibre, {\it Integrability of polynomial differential systems, Handbook of
Differential equations, Ordinary Differential Equations}, volume 1, Chapter
5, pages 437-531. Edited by A. Ca\~nada, P. Dr\'abek and A. Fonda. Elsevier
B.V. 2004.

\bibitem{firsTHEOps1} L.G.S. Duarte, S.E.S.Duarte and L.A.C.P. da Mota,
{\it A method to tackle first order ordinary differential equations with
Liouvillian functions in the solution}, in {\it J. Phys. A: Math. Gen.}
{\bf 35} 3899-3910 (2002).

\bibitem{secondTHEOps1} L.G.S. Duarte, S.E.S.Duarte and L.A.C.P. da Mota,
{\it Analyzing the Structure of the Integrating Factors for First Order
Ordinary Differential Equations with Liouvillian Functions in the
Solution}, {\it J. Phys. A: Math. Gen.} {\bf 35} 1001-1006 (2002)

\bibitem{nossoPS1CPC}  L.G.S. Duarte, S.E.S.Duarte, L.A.C.P. da Mota and J.F.E.
Skea, {\it Extension of the Prelle-Singer Method and a MAPLE
implementation}, {\it Computer Physics Communications}, Holanda, v. 144, n. 1, p. 46-62, 2002


\bibitem{davenport}
Davenport J.H., Siret Y. and Tournier E.
{\it Computer Algebra: Systems and Algorithms for Algebraic Computation}.
Academic Press, Great Britain (1993).

\bibitem{ManMac}
Y K Man and M A H MacCallum,
A Rational Approach to the Prelle-Singer Algorithm.
{\it J. Symbolic Computation}, {\bf  11} 1--11 (1996), and refferences therein.


\bibitem{PS2}
L G S Duarte, S E S Duarte, L A C P da Mota and J E F Skea,
Solving second order ordinary differential equations by extending the Prelle-Singer method,
{\it J. Phys. A: Math.Gen.}, {\bf 34} 3015-3024 (2001).

\bibitem{royal}
Chandrasekar, V K ; Senthilvelan, M ; Lakshmanan, M.,
On the complete integrability and linearization of certain second
order nonlinear ordinary differential equations.
{\it http://arxiv.org/abs/nlin/0408053 - accepted for publication on
Proceedings of the Royal Society London Series A}

\bibitem{buchdahl} Buchdahl, H.A.,
Relativistic fluid spheres resembling the Emden polytrope of index
5.
{\it Ap. J. 140}, 1512-1516. (1964).

\bibitem{Almendal}
Almendral, J. A. and Sanju´an, M. A. F.,
Integrability and symmetries for the Helmholtz
oscillator with friction.
{\it J. Phys. A36}, 695-710 (2003)

\bibitem{LakshmananRajasekar}
Lakshmanan, M. and Rajasekar, S.,
{\it Nonlinear Dynamics: Integrability, Chaos and Pat-
terns.}. New York: Springer-Verlag (2003).

\end{thebibliography}
\end{document}